# Electron Transport in Compacted Powders of VO₂ Nanoparticles: Variable Range Hopping vs Percolation Behavior

E.Yu. Beliayev[1,2*], Yogendra Kumar Mishra[2], Horst-Günter Rubahn[2], I.A. Chichibaba[3], I.G. Mirzoiev[1], V.A. Horielyi[1], A.V. Terekhov[1]

[1]B. Verkin Institute for Low Temperature Physics and Engineering of the National Academy of Sciences of Ukraine

[2]Smart Materials Group, Mads Clausen Institute, University of Southern Denmark, Sønderborg.

[3]National Technical University «Kharkiv Polytechnic Institute», Kharkiv, Ukraine.

e-mail*: beliayev@ilt.kharkov.ua

Electron transport properties in compacted VO₂ nanopowders were studied. While VO₂ usually exhibits a first-order metal-insulator transition (MIT) at ~340 K, in our compressed nanopowder samples MIT was significantly broadened due to structural disorder, interparticle barriers, and phase coexistence. Resistivity measurements in the temperature range of 78–682 K initially suggested a Variable Range Hopping transport mechanism, but further analysis indicates that the observed temperature dependence is governed by percolative conductivity, modified by activation-assisted tunneling effects. Suppression of the expected resistance jump at the MIT is attributed to dynamic intergranular barrier restructuring, residual localized states, and percolative electron transport. These findings highlight the necessity of considering percolation effects when analyzing transport mechanisms in granular VO₂-based systems.

**Introduction**

Vanadium dioxide is a well-known transition metal oxide that exhibits a metal-insulator transition (MIT) around the temperature of 340 K. The phase transition temperature can be changed by introducing certain dopants [1] or varying nanoparticle morphology [2] [3]. The MIT is usually accompanied by a sharp change in conductivity, typically by several orders of magnitude. In its high-temperature metallic phase, VO₂ adopts a rutile-type structure characterized by a high density of states (DOS) at the Fermi level. Conversely, in its low-temperature insulating phase, VO₂ transitions to a monoclinic (P21/c) structure and exhibits an antiferromagnetic ordering, accompanied by a significant reduction in both the DOS and electrical conductivity ($\sigma$).

Vanadium dioxide has four most common polymorphs: VO₂ (A) (tetragonal), VO₂ (B) (high-temperature monoclinic phase with metallic type conductivity), VO₂ (R) (tetragonal-rutile), and VO₂ (M) (monoclinic-distorted rutile) [4] and their relative stability can be strongly influenced by external factors, such as nanoparticle morphology, internal stress, and pressure [5]. Additionally, VO₂ demonstrates an almost unlimited variability of valence states within a homologous series of thermodynamically stable and metastable vanadium oxides. These oxides belong to the so-called Magnéli ($V_nO_{2n-1}$) and Wadsley ($V_{2n}O_{5n-2}$) series of vanadium–oxygen phases [6], including but not limited to $V_2O_3$, $V_3O_5$, $VO_2$, $V_3O_7$, $V_2O_5$, $V_2O_2$, $V_6O_{13}$, and $V_4O_9$ [7]). This structural and chemical diversity makes it a very complex object of research.

Although the electron transport properties of VO₂ have been extensively studied, it has been reported that at nanoscopic dimensions, these properties change drastically [8]. Despite the vast amount of research available in the literature, many aspects of VO₂ conductivity in its nanoparticle form remain poorly understood, particularly in compressed VO₂ nanopowders, where pressing introduces additional structural disorder and influences interfacial conductivity, thereby leading to enhanced electron localization. The aim of this work has been to investigate the electrical transport properties of compressed VO₂ nanopowders, focusing on their behavior near the metal-insulator transition. A particularly intriguing aspect is whether a reversible metal-insulator transition can be induced in the granular VO₂-based systems under an applied electric field—an effect reminiscent of that observed in ultrathin percolating gold films [9]. Given the growing use of VO₂ nanopowders in functional materials and electronic devices, understanding their transport mechanisms, including the effects of structural disorder and phase coexistence on the metal-insulator transition, is of significant practical importance.

**Samples and Experimental Methods**

In this work, we studied the compressed nanopowders consisting of $VO_2$ nanoparticles obtained via the hydrothermal synthesis method (Fig. 1). During synthesis, a 1:1 mixture of $V_2O_5$ as a vanadium source and 98% oxalic acid ($H_2C_2O_4 \cdot 2H_2O$) were used as the reducing agent followed by the addition of distilled water. The mixture was heated to 60 °C with stirring until a clear dark blue solution was formed, and 10 mol % of tartaric acid was added as a complexing agent for vanadium ions. Then, the mixture was kept in an autoclave at 240 °C for 12 h. After completion of the reaction and cooling to room temperature, the resulting black precipitate was carefully collected by centrifugation, washed with distilled water and ethanol, and dried in air at $T = 80$ °C for 12 h. A similar approach was previously applied in [10] for the synthesis of $CrO_2$ nanoparticles, demonstrating its effectiveness in producing rutile-type oxide materials.

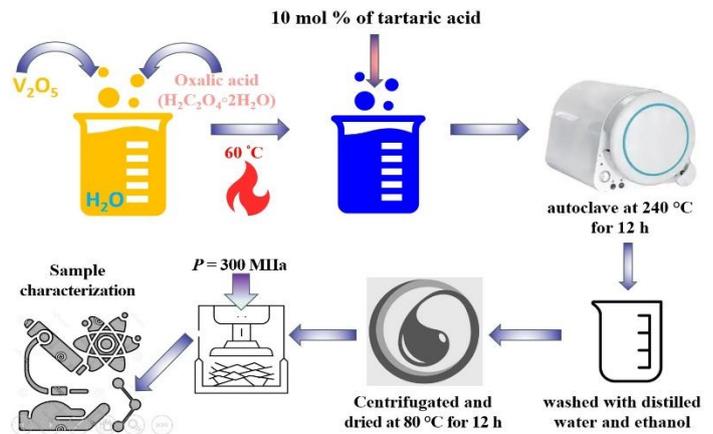

Fig. 1. Steps involved in the hydrothermal synthesis of $VO_2$ nanoparticles: The black $VO_2$ nanopowder was compressed into round pellets and later cut into rectangular parallelepiped shapes for transport studies using a slow-rotating diamond cutter to prevent overheating and possible decomposition.

The morphology of the synthesized $VO_2$ nanoparticles was studied by a Scanning Electron Microscope (Hitachi S3400N) at an accelerating voltage of 10 kV with a working distance of 10 mm. The analysis revealed that the nanoparticles synthesized hydrothermally were a kind of rod-shaped, with a thickness/diameter of around 70 to 100 nm and lengths up to 1 μm (Fig. 2).

According to *X*-ray phase analysis carried out with a Rigaku MiniFlex II instrument with Cu-$K_\alpha$ monochromatic radiation ($\lambda = 1.54056$ Å), the initial samples contained 99% of two monoclinic modifications of vanadium dioxide: $VO_2$ (A) with space group P42/*ncm* (PDF 01-070-2716) and $VO_2$ (B) with space group C2/*m* (PDF 01-081-2392), which is typical for hydrothermally obtained $VO_2$ nanopowders. The coexistence of these phases suggests potential structural inhomogeneities that could contribute to the broadening of the metal-insulator transition.

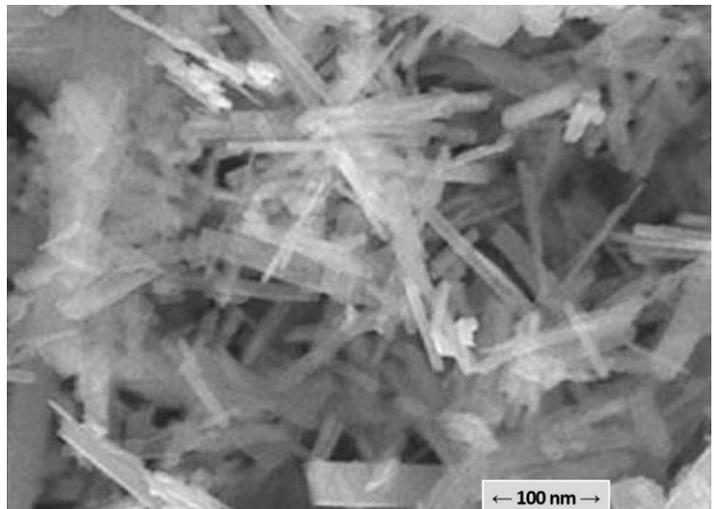

Fig. 2. Typical SEM images of $VO_2$ nanoparticles synthesized by the hydrothermal method.

*Low-temperature measurements*

All the electrical measurements were made in the Department of Transport Properties of Conducting and Superconducting Systems at B. Verkin Institute for Low Temperature Physics and Engineering of the National Academy of Sciences of Ukraine (Kharkiv), in the Laboratory of Electronic, Magnetic, and Superconducting Properties of Metals and Multicomponent Compounds.

The samples for measurements were made by cold pressing at 30 MPa. The sample was shaped in regular parallelepiped form with $11 \times 2 \times 1.5$ mm³ dimensions for electrical transport measurements. On the upper edge of the sample, contact pads were formed by vacuum deposition of silver, to which the measuring wires were attached using conductive silver paste. The distance in between the potential contacts was around 6 mm. Low-temperature measurements were made in a vacuum cryostat in the temperature range of 78 K - 400 K by a standard four-probe direct current method with an applied current of 100 μA. The current was measured using a Keithley 2000 multimeter and reversed at each temperature to eliminate the influence of thermoelectric EMF (electromotive force), and a Keithley 2182 nanovoltmeter controlled the measuring voltage. The temperature at each measurement point was controlled by a custom-made PID controller with an accuracy of 0.005 K, limiting temperature fluctuations that could lead to transition blurring.

Although the resistance changed by six orders of magnitude in the temperature range of 78 K - 400 K, no noticeable features were found on $R(T)$ dependence near the usually observed transition temperature $T = 340$ K for our sample at first glance (see Fig. 5). Applying a magnetic field up to $B = 1.7$ T did not alter the sample's resistance and the current-voltage (I-V) characteristics measured at $T = 78, 148, 201, 292,$ and 392 K (Fig. 3) also did not shed light on the situation. The I-V characteristics were recorded for both voltage polarities to account for potential thermoelectric effects. The data presented in Fig. 3 represent the averaged values, ensuring that any asymmetry or possible thermoelectric contributions are minimized.

Despite the detailed control of sample preparation and precise measurement setup, the resistance measurements showed no sharp resistive changes, pertinent to MIT up to the maximum possible measuring temperature for the cryostat, which was 400 K. This motivated us to extend the measurement range to higher temperatures.

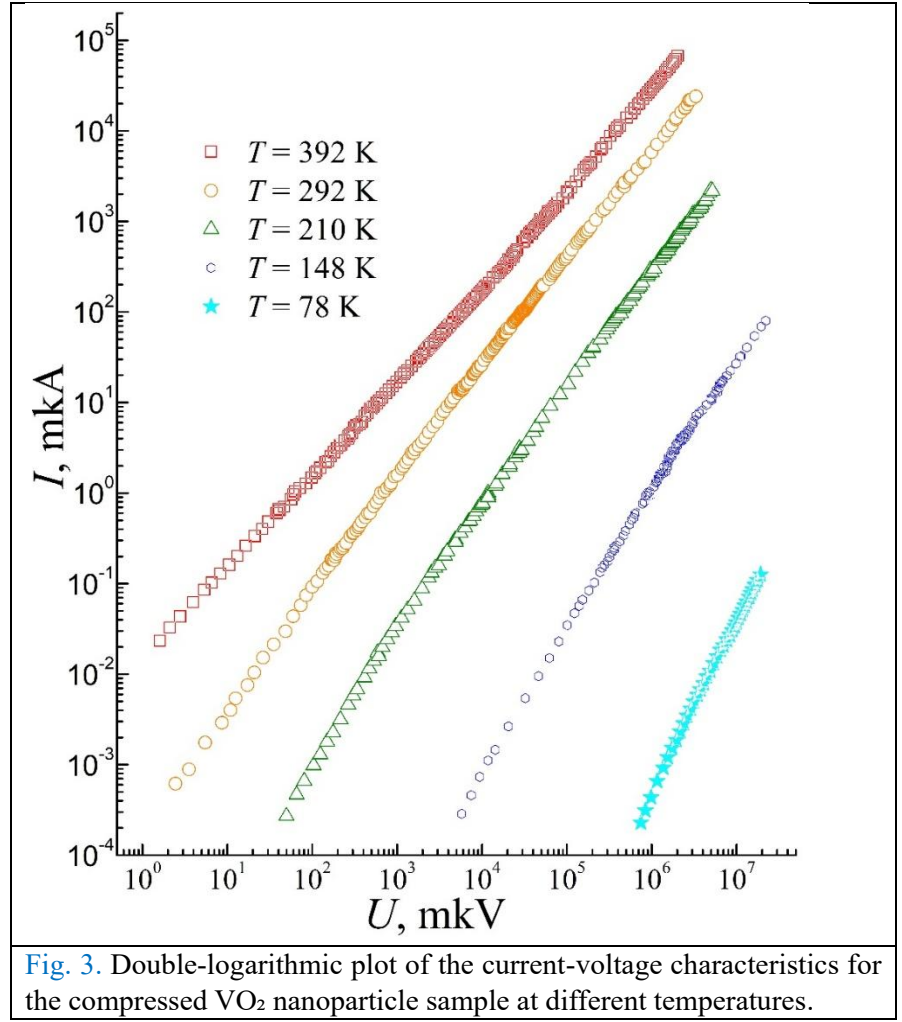

Fig. 3. Double-logarithmic plot of the current-voltage characteristics for the compressed $VO_2$ nanoparticle sample at different temperatures.

### High-temperature measurements

A special thermostat (Fig. 4) was designed for high-temperature measurements. The sample holder consisted of copper and sitall (a glass-ceramic material) plates, with a heating element made of thin nichrome wire placed between them, separated by a layer of mica. The copper plate was equipped with a groove for a calibrated chromel-alumel thermocouple to measure and maintain temperature accurately. The sample was cleaned of its original contacts, and identical contacts were made on the opposite side by vacuum deposition of 99.99% pure gold using the same mask. The gold layer had a thickness of 0.5 μm to eliminate the possibility that the blurring of the transition is due to poor contacts. Four molybdenum spring contacts pressed the measured sample against a ceramic plate. For reliable electrical contact, gold wires were tightly wound, turn by turn, around the ends of the molybdenum springs. These gold wires made contact with the gold-plated pads on the sample. To ensure insulation, the ends of the gold wires were passed through thin fluoroplastic tubes and routed out of the hot zone. Once outside the hot zone, the wires were soldered to an electrical connector, which allowed them to be connected to the measuring equipment. The measuring cell was wrapped in thin mineral wool to prevent convection and covered with a copper shield to minimize radiative heat exchange. The device allowed measurements to be carried out in the temperature range of 300 K – 720 K using the same equipment and methods as were previously used in the cryostat.

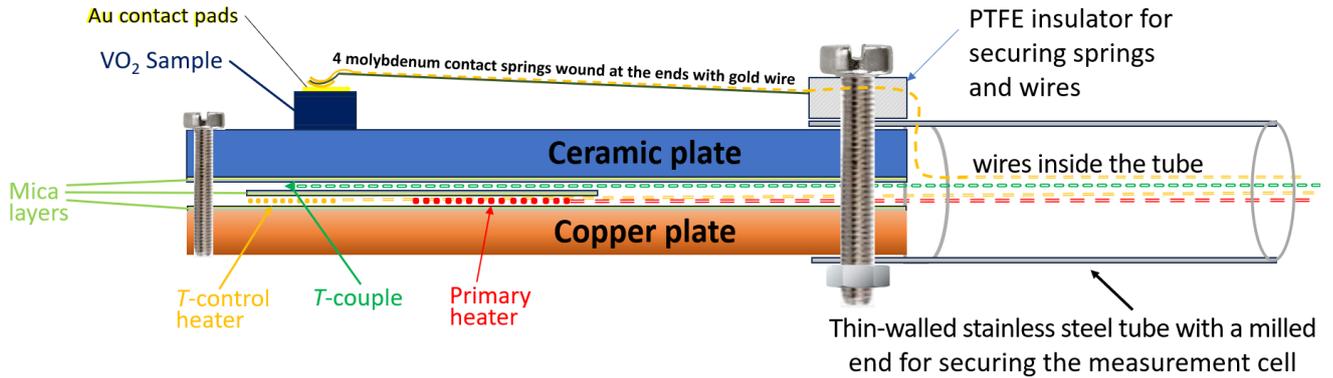

Fig. 4. High-temperature thermally controlled measurement setup for VO₂ samples.

**Results and discussion**

The $R(T)$ dependence obtained from the combined 'low-temperature' and 'high-temperature' measurements taken at measuring current $I = 100$ mkA is shown in Fig. 5. This curve provides key data for identifying the dominant conduction mechanism and its relation to the metal-insulator transition.

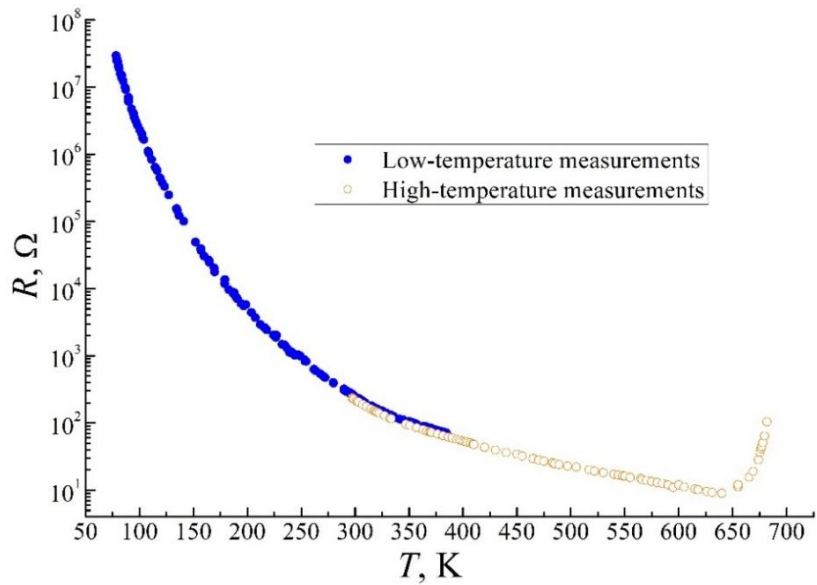

Fig. 5. Temperature-dependent resistance of pressed VO₂ nanoparticle sample ($T = 78 – 682$ K), measured at $I = 50$ mkA.

The overall temperature dependence of the sample resistance looked quite smooth. When measured in the opposite direction (from high to low temperatures), a small thermal hysteresis could be observed. However, such hysteresis was observed in almost all temperature ranges, and it was difficult to separate it from the overall hysteresis of the cryostat.

When the measurement temperature exceeded $T = 640$ K, the sample's resistance increased sharply, indicating the onset of thermal degradation in vanadium dioxide. This was confirmed by subsequent *X*-ray analysis, which revealed irreversible thermal decomposition of the sample.

The most studied phases associated with the metal-insulator transition (MIT) are VO₂ (R) (rutile) and VO₂ (M) (monoclinic), with the transition typically occurring around $T = 340$ K. It is known from the available literature that VO₂ powder samples often exhibit smoother phase transitions and smaller conductivity jumps compared to single crystals, due to polymorphism [2] induced during pressing, structural inhomogeneities, and localized mechanical stresses at grain boundaries [11].

The coexistence of $VO_2$ (R) and $VO_2$ (M) phases, combined with pressure-induced stabilization of intermediate phases, contributes to the smearing of the transition by creating regions with different local transition temperatures and possible ≈ 1 % of $V_2O_5$ admixture forming thin dielectric barriers across grain boundaries. At the same time the stability of phase composition in $VO_2$ is highly sensitive to pressure, with moderate changes capable of shifting the transition temperature or inducing metastable states [12]. Local stress and strain at grain boundaries in pressed powders may shift or suppress the MIT. Additionally, oxygen non-stoichiometry and grain interface features [11] may alter local electronic properties, further diffusing the transition's sharpness.

This explains the absence of a sharp conductivity jump on the *R(T)* curve, suggesting that the dominant role of structural and morphological factors, such as phase coexistence and interparticle barriers, diffuses the transition, while the sample's percolative nature further masks its intrinsic effects.

Although the general curve shown in Fig. 5, obtained from the combined results of low- and high-temperature measurements, seems pretty smooth, two linear regions could be distinguished on the electrical resistance versus temperature plotted in $ln(R) = f(T^{-1/4})$ coordinates (see Fig. 6) presumably corresponding to the manifestation of the variable-range hopping (VRH) conductivity mechanism in a three-dimensional case described by the formula:

$$\rho(T) = \rho_0 \times exp\left[\left(\frac{T_M}{T}\right)^{1/4}\right], \quad (1)$$

where $T_M$ is the Mott characteristic temperature, which depends on the density of localized states at the Fermi level $N(E_F)$ and the localization length ξ, while $\rho_0$ is a material-dependent pre-exponential factor, typically on the order of Ohms or higher for materials with structural disorder.

For the experimentally measured *R(T)* dependence, plotted in Mott's coordinates, the Mott temperature $T_M$ is related to the slope through the following equation:

$$slope = \left(\frac{T_M}{T}\right)^{1/4} \quad (2)$$

Fig. 6 illustrates the temperature dependence of the sample's resistance alongside the fitted curves corresponding to the 3D case of Mott VRH theory in both the low-temperature and high-temperature measurement ranges. The associated Mott temperatures and the quality of the fit, represented by dashed lines, are also depicted. Notably, the inflection point in the *R(T)* curve in Mott coordinates and the intersection of the fitted lines provide clear evidence of the phase transition and metal-insulator transition around *T* = 340 K, pertinent to most vanadium dioxide samples as it is generally known that vanadium dioxide ($VO_2$) undergoes a first-order metal-insulator transition (MIT) at approximately 340 K, shifting from a semiconducting monoclinic (M1) phase to a metallic rutile (R) phase. This transition has been a subject of significant interest since its discovery in the late 1950s [13].

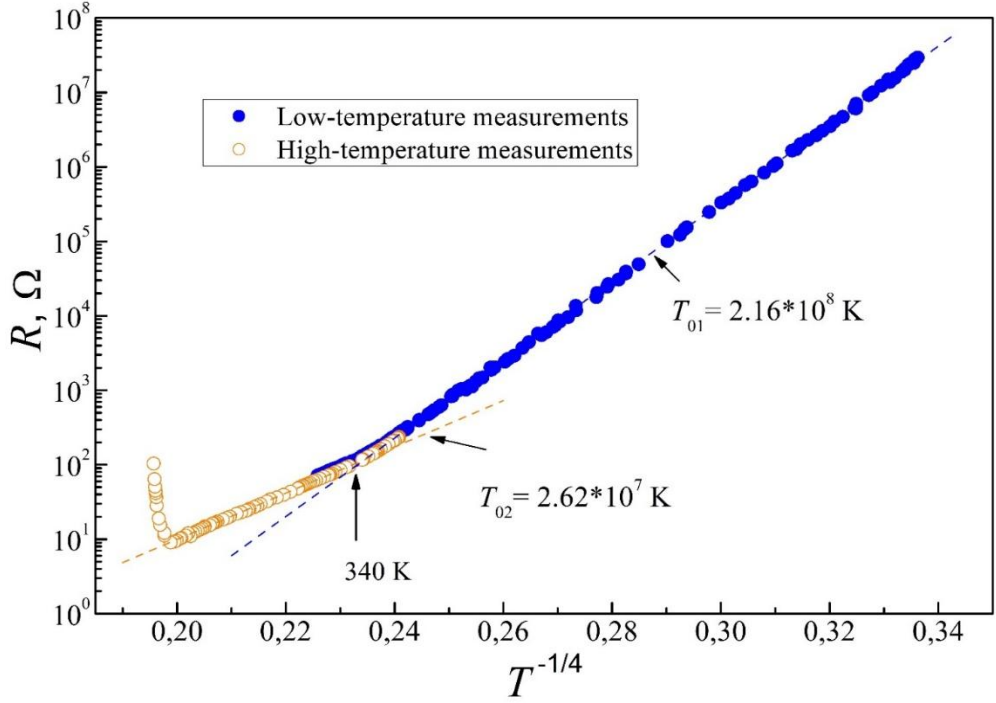

Fig. 6. Resistance of VO$_2$ pressed powder sample plotted as $lnR$ versus $T^{-1/4}$.

Attempts to fit the experimental results to other conductivity mechanisms have been unsuccessful. Intuitively, the manifestation of the 3D hopping conduction mechanism in a granular medium, such as the pressed VO$_2$ nanoparticle powder, seems reasonable since structural disorder and the presence of potential barriers between particles create conditions for electron localization. The fact that $ln(R) \propto T^{-1/4}$ law held over more than six orders of magnitude in the sample's resistance in the dielectric (narrow gap semiconductor with $E_g \approx 0.6 – 0.7$ eV) state of the VO$_2$ sample strongly suggested that we were observing three-dimensional Mott hopping conductivity and warranted further analysis in this direction. Moreover, the reduction in the slope of the $ln(R) = f(T^{-1/4})$ dependence at $T = 340$ K was in line with a sudden increase in the density of electronic states at the Fermi level during the metal-insulator transition, and thus we delve into the analysis.

If the density of states $N(E_F)$ is known, we can use the Mott temperature $T_M$, previously calculated from Fig. 6, to determine the localization length ξ in VO$_2$ for both metallic and insulating phases. The Mott temperature, $T_M$, is related to the localization length ξ through the following formulae [14]:

$$T_M = \frac{\beta}{k_B N(E_F) \cdot \xi^3} \Leftrightarrow \xi = \left(\frac{\beta}{k_B T \cdot N(E_F) \cdot T_M}\right)^{1/3}. \qquad (3)$$

Here, β is a numerical coefficient, which depends on the system's dimensionality (for a 3D system, $\beta \approx 21 \pm 1.2$, for 2D case $\beta \approx 13.8 \pm 0.8$ [14]), while ξ represents the localization length.

Combining the density of states $N(E_F)$, either calculated or experimentally measured, with the Mott temperature $T_M$ determined from Fig. 6, it would be interesting to calculate the localization length ξ, providing deeper insight into the metal-insulator phase transition in VO$_2$. However, a major obstacle emerged: obtaining reliable data for the density of electronic states at the Fermi level for vanadium dioxide.

Vanadium dioxide has been the subject of a huge number of articles, including reviews, devoted to the topic of the metal-insulator transition [15][16][17]. However, after reviewing the available literature, it appears that there are no readily accessible numerical values for the density of states $N(E_F)$ for VO$_2$ directly supported by DFT (Density Functional Theory) calculations or other reliable methods. While many studies focus on the band structure and density of states for VO$_2$ [18], the specific $N(E_F)$ values under DFT theory seem unreported or buried within broader discussions about the material's electronic structure. For example, [19] discusses the electronic structure of different phases of VO$_2$, focusing on orbital contributions and band structure near the Fermi level, but does not provide explicit values for $N(E_F)$. Similarly, [20] provides the comparative DOS calculations for metallic and insulating $VO_2$ phases. Still, specific numerical data for $N(E_F)$ remained elusive.

Another reliable data source for estimating $N(E_F)$ for $VO_2$ could be photoemission techniques such as PES and ARPES measurements. These methods have been widely applied to study $VO_2$'s electronic structure, but direct $N(E_F)$ values are generally absent. For instance, [21] discusses ARPES measurements in quantum materials, including $VO_2$, but specific $N(E_F)$ values in $VO_2$ are not given. Also, the paper [22] does not provide explicit numerical values for $N(E_F)$ in $VO_2$. Instead, it discusses the ARPES spectra showing a dispersive feature of the $O_{2p}$ band and the periodicity of the dispersive bands, which are consistent with the expected structure of $VO_2$ in the metallic phase, providing insight into the behavior of the electronic states around the metal-insulator transition. Moreover, ARPES results often reflect surface-sensitive properties, which may not accurately represent bulk electronic states due to vanadium's surface valence variability.

Since vanadium dioxide is generally recognized as a strongly correlated material [23], any analysis of voltage-current characteristics following the relation $dI/dV \propto N(E_F)$ to estimate the density of states at the Fermi level $N(E_F)$ is inappropriate. When applied to the CVC shown in Fig. 7, this approach yields a non-physical value $N(E_F) \approx 1.23 \times 10^{20}$ eV$^{-1}$·m$^{-3}$ $\approx 7.7 \times 10^{38}$ J$^{-1}$·m$^{-3}$, as it fails to account for the complex electron interactions characteristic of strongly correlated systems, which greatly enhance the density of states at the Fermi level. As a result, this value is significantly underestimated compared to, for example, the data for $CrO_2$ (chromium, a neighbor of vanadium in the periodic table) reported in the known literature [24], $N(E_F) \approx 1.51 \times 10^{47}$ J$^{-1}$·m$^{-3}$. Therefore, more reliable calculation techniques ought to be used.

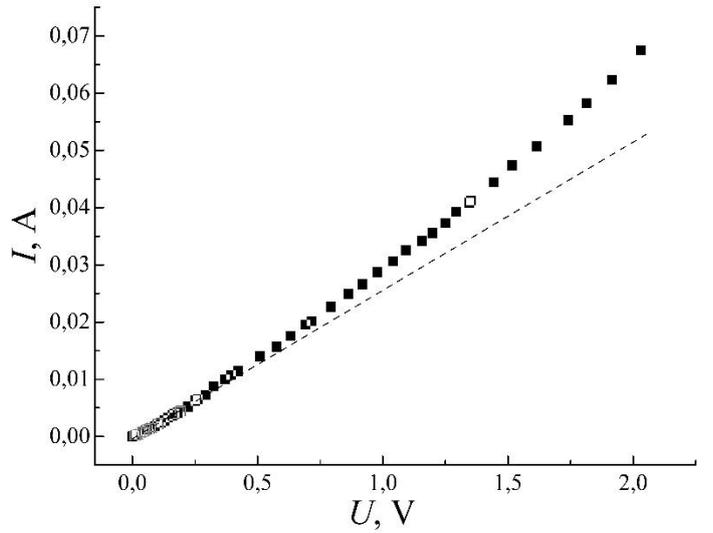

Fig. 7. Voltage-Current Characteristic for $VO_2$ pressed powder sample, measured in metallic state at $T = 392$ K. Dashed line shows the linear fit at low voltages.

Another possible method for estimating $N(E_F)$ involves the electronic contribution to the heat capacity $C_{elec}$, where the heat capacity coefficient $\gamma$ is related to the density of states by the equation

$$\gamma = \frac{\pi^2 k_B^2}{3} N(E_F). \tag{4}$$

Studies on the heat capacity of transition metal oxides [16] [17] provide data for $VO_2$ in its metallic phase. However, the data in these studies are presented in an unseparated form, whereas Equation (3) requires the isolated electronic component of the heat capacity.

Attempt to estimate $N(E_F)$ using the total heat capacity results in a non-physical value $N(E_F) \approx 7.08 \times 10^{44}$ state/eV·m$^{-3}$ $\approx 4.42 \times 10^{63}$ state/J$^{-1}$·m$^{-3}$ for $VO_2$. This discrepancy arises because the phonon contribution to the heat capacity dominates at temperatures above the MIT point (e.g., $T = 392$ K), and the electronic contribution becomes negligible at low temperatures when $VO_2$ is in its insulating state. Additionally, the Fermi-liquid model, which underlies Equation (3), is not applicable to $d$-metals with strong electron correlations, further invalidating this approach for vanadium dioxide. Reports of anomalously low electronic thermal conductivity in the metallic state of $VO_2$, which violates the Wiedemann-Franz law [25], especially near the metal-insulator transition, further highlight the limitations of this method.

Unlike other strongly correlated systems, such as $La_{2-x}Sr_xCuO_4$, where antiferromagnetic order strongly influences transport properties, $VO_2$ does not exhibit long-range magnetic ordering. Instead, its metal-insulator transition is primarily driven by structural distortions and electron-lattice interactions. This distinction is important when analyzing electronic states, as standard magnetic susceptibility measurements, $\chi$, may not reliably reflect the density of states at the Fermi level, as discussed in [26]. Furthermore, susceptibility values often include unseparated diamagnetic ($\chi_d$) and orbital ($\chi_{orb}$) contributions, which are independent of conduction electrons, making their direct use for estimating $N(E_F)$ problematic. The orbital susceptibility, $\chi_{orb}$, arises from electron motion within atomic orbitals and is independent of the density of states at the Fermi level. Instead, it originates from bound electron clouds that do not participate in conduction. Similarly, the diamagnetic contribution, $\chi_d$, reflects the response of all filled orbitals (valence and core) and is an inherent property of any closed-shell system, unrelated to the availability of conduction electrons.

At last, in the ongoing attempts to find or calculate the DOS value $N(E_F)$, we come across the review article [27], in which Jean-Paul Pouget refers to his earlier works [28] and [29]. There, the effective density of states $N_{eff}(E_F)$ could be derived from magnetic susceptibility measurements in the metallic phase of the $VO_2$ powder sample (at $T = 370$ K), combined with the NMR measurements on $^{51}V$ nuclei. After isolating the orbital contribution, the authors analyzed the relationship between the Knight shift $K$ and the total magnetic susceptibility $\chi$, further separating the spin component $\chi_s$ by subtracting the diamagnetic contribution $\chi_d = 6.2 \times 10^{-4}$ emu/mole $VO_2$ (at $T = 370$ K). This analysis yields the effective density of states at the Fermi level $N_{eff}(E_F) \approx 10$ state/(eV × formula unit × spin direction).

The spin susceptibility $\chi_s$ for the $VO_2$ powder sample was deduced from the plot of the isotropic part of the $^{51}V$ NMR Knight shift tensor $K$ by decomposing the total magnetic susceptibility into its diamagnetic, orbital, and spin contributions, where the spin magnetic susceptibility $\chi_s$, is directly related to the density of states at the Fermi level by a simple relation $\chi_s = \mu_B^2 \times N_{eff}(E_F)$ within the framework of Pauli's theory of paramagnetism.

It is important to highlight that the value $N_{eff}(E_F)$ obtained by this method represents an *effective density of states* at the Fermi level, incorporating electronic correlations within the material. This differs from the more straightforward $N(E_F)$ derived from free-electron models. For strongly correlated systems like $VO_2$, electron-electron interactions significantly enhance the actual density of states, making $N_{eff}(E_F)$ much larger than the values predicted by simplified approaches, such as the Local Density Approximation (LDA). Similar effects have been observed in $RM_4Al_8$ compounds with the $ThMn_{12}$-type crystal structure, where complex interactions, including Kondo effects and spin-glass states, significantly influence electronic transport [30]. Theoretical LDA-based values typically yield only 1.24–1.31 states/eV per $VO_2$ formula unit per spin direction, while the experimentally measured $N_{eff}(E_F)$ reflects the true impact of $VO_2$ electronic correlations.

Recent studies [31] have confirmed Jean-Paul Pouget's results, extending them to include insights into anisotropic susceptibility and orbital effects, providing additional information on the electronic structure of $VO_2$ and complementing earlier analyses [27], [28] and [29], but without changing their essence.

So, we have $N_{eff}(E_F) \approx 2 \times 10 = 20$ state/(eV×formula unit) for two spin directions. To convert the density of states per formula unit to density per cubic meter, we need to determine how many $VO_2$ formula units are contained in one cubic meter. The molecular mass of $VO_2$ is $M_{VO2} = 83.996$ g/mol, and the density of $VO_2$ is $\rho = 4.6$ g/cm$^3$ = $4.6 \times 10^6$ g/m$^3$. Thus, one cubic meter contains $\rho/M_{VO2} = (4.6 \times 10^6$ g·m$^{-3})/(83.996$ g·mol$^{-1}) \approx 5.48 \times 10^4$ mol·m$^{-3}$. To find the number of formula units per cubic meter, we multiply this value by Avogadro's number: $5.48 \times 10^4$ mol·m$^{-3}$ × $6.022 \times 10^{23} \approx 3.3 \times 10^{28}$ formula units/m$^3$. Now, multiplying by the density of states per formula unit, we get $N_{eff}(E_F) = 20$ states/(eV × formula unit) × $3.3 \times 10^{28}$ formula units/m$^3 \approx 6.68 \times 10^{29}$ states/(eV·m$^3$). Then, using 1 eV=$1.60218 \times 10^{-19}$ J, we find $N_{eff}(E_F) = 6.68 \times 10^{29}$ states/(eV·m$^3$)/$1.60218 \times 10^{-19}$ J/eV $\approx 4.17 \times 10^{48}$ states/J·m$^{-3}$.

- The method for estimating $N_{eff}(E_F)$ value for $VO_2$ through the spin component of magnetic susceptibility, as determined by Knight shift measurements by Jean-Paul Pouget and co-workers, is reliable. The comparison with the value $N(E_F) \approx 1.51 \times 10^{47}$ J$^{-1}$·m$^{-3}$ for $CrO_2$ in the metallic phase, taken from the known literature [24], confirms its correctness. Therefore, for $VO_2$ in the metallic phase, we take $N_{eff}(E_F) \approx 6.68 \times 10^{29}$ states/eV·m$^{-3}$ = $4.17 \times 10^{48}$ states/J·m$^{-3}$.
- The values of $T_M = 2.16 \times 10^8$ K for the insulating phase and $T_M = 2.62 \times 10^7$ K for metallic phases (estimated from the $ln(R) = f(T^{-1/4})$ plot Fig. 6) can be accepted as an experimental fact. Due to Peierls cell doubling during the metal-insulator transition, the density of states for $VO_2$ is expected to decrease by half in the insulating phase.

However, substituting the obtained values into Equation (3) yields a non-physical localization radius $\xi \approx 1.1168 \times 10^{-11}$ m, which turns out to be less than the diameter of an atom. Such a result indicates the breakdown of the physical assumptions underlying the Mott variable range hopping (VRH) model, at least at high temperatures, for our $VO_2$ powder system. At lower temperatures, however, where the density of localized states is reduced, the VRH mechanism may still be relevant.

Despite the apparent good fit of $R(T)$ to the Mott VRH model, shown in Fig. 6, the small localization radius indicates strong electron correlations that further restrict carrier mobility, even in the metallic phase. Also, the Mott model, which assumes phonon-assisted hopping between localized states, is typically valid only at low temperatures (up to $T = 0.2 \div 0.3 \times \Theta$ Debye), while for $VO_2$, $\Theta(T) \approx 600 \div 750$ K [32]. At higher temperatures, this model loses applicability, requiring theories such as the Hubbard model or DFT+U.

The Peierls-like mechanism of unit cell doubling during the metal-insulator transition suggests that lattice restructuring plays a major role in the observed resistance behavior. This restructuring affects the geometry and connectivity of $VO_2$ grains, modifying current pathways, enhancing the percolative nature of conduction, and contributing to the change of slope on the Mott plot (Fig. 6). In the insulating phase, the reduced density of states at the Fermi level further suppresses conduction.

It is worth mentioning that in many known studies [33][34][35], even for high-quality films and well-formed crystalline samples of vanadium dioxide after the insulator-metal transition, there is often no resistive minimum and no change in the sign of $dR/dT$, which is considered to be an attribute of a completed transition to the metallic state. And our study of the $VO_2$ $R(T)$ dependence, carried out in the widest possible temperature range, we also did not reveal a minimum, which suggests that $VO_2$ retains some localized states even in its metallic phase. This may result from thermal fluctuations or incomplete phase transitions when domains of the insulating phase coexist with metallic regions. Structural defects or instabilities in the metallic phase could further reinforce this mixed-phase scenario. So even in the metallic state, insulating domains may persist, contributing to a mixture of localized and delocalized electronic transport.

In the high-temperature domain (340 – 620 K), limited resistance changes visible on the Mott plot (Fig. 6) suggest that thermal fluctuations dominate, potentially masking the expected metallic behavior. The increasing electron mobility in the metallic phase with temperature remains indistinguishable, partly due to a narrow one-decade resistance range observed in this region.

The lack of a clear resistive minimum—typical of metallic behavior—implies that even after the transition, $VO_2$ retains localized states, leading to non-metallic properties. This persistence of localized states is often linked to structural features such as defects or grain boundaries, which promote a mixed-phase scenario where metallic and insulating domains coexist. The small localization radius and structural defects may further reinforce this phase mixing, resulting in anomalous resistance behavior.

Additionally, experiments with thin films and $VO_2$ nanoparticles [36] show that resistance noise and thermal fluctuations can significantly influence the resistance behavior, particularly in unstable structures or regions undergoing phase transitions.

Also, temperature fluctuations associated with the PID-controlled system during high-temperature measurements could introduce further variability, complicating the analysis of resistance behavior and highlighting the sensitivity of $VO_2$ to external conditions.

Meanwhile, the observed formal agreement of the temperature-dependent resistance behavior with Mott's law, as seen in Fig. 6, apparently requires further explanation.

At this point, we must consider the sample's structure—a pressed metal powder in which the measuring current is unevenly distributed due to the granular morphology and the varying heights of the contact barriers between adjacent $VO_2$ nanoparticles. *X*-ray diffraction reveals the presence of approximately 1% of other vanadium oxides despite the high claimed purity. The presence of residual $V_2O_5$ on the surfaces of $VO_2$ grains, likely originating from the synthesis process or subsequent environmental oxidation, further enhances the percolative nature by creating insulating barriers that influence the tunneling probability and intergrain connectivity.

Since $V_2O_5$ is known to be a semiconductor or insulator, it likely forms surface layers on the $VO_2$ particles, creating energy barriers between conductive $VO_2$ grains. These barriers enhance the percolative nature of sample's conductivity, as electrons must traverse insulating interfaces.

In a granular $VO_2$ system, whose conductivity is dominated by barriers caused by surface impurities such as $V_2O_5$ or by incomplete particle connectivity, the conductivity follows percolation mechanisms. Near the percolation threshold, the effective resistance change associated with the metal-insulator phase transition can be significantly suppressed due to incomplete connectivity and the presence of insulating barriers between conducting regions, compared to well-connected systems. This reduction arises because the resistance change depends not only on the intrinsic properties of $VO_2$ but also on the intergrain connectivity and the distribution of barrier heights. The percolative nature of the sample likely explains the less pronounced resistance change observed during the MIT compared to single-crystal or highly connected $VO_2$ systems.

Furthermore, the needle-like shape of the $VO_2$ nanorods (aspect ratio ~1:10) introduces anisotropy in the percolation pathways, further suppressing the resistance jump at the phase transition threshold. Thus, the sample effectively behaves as a three-dimensional percolative conductive network, where the metal-insulator phase transition occurs within a certain temperature range near $T = 340$ K.

### Interplay of Percolation and VRH in Electrical Conductivity of Pressed VO₂ Powder Samples

The initial interpretation of the resistance data as being governed by Mott's variable-range hopping (VRH) conductivity stemmed from the strong agreement between the experimentally measured temperature dependence and the expected VRH dependence (Fig. 6), particularly in the low-temperature range. The resistance, plotted in Mott coordinates as $ln(R) = f(T^{-1/4})$, displayed two distinct linear regions with an inflection point near the metal-insulator transition at $T = 340$ K. However, further investigation, considering the density of states near the Fermi level pertinent to $VO_2$, as determined from Knight shift data [27], showed that the localization length ξ associated with the Mott variable-range hopping mechanism, given by Formula 3, is unphysically small. From this, we concluded that the observed match between the experimental data and VRH behavior is not a manifestation of true Mott hopping itself but is, in fact, a result of the complex interplay between percolative conductivity mechanisms and electron tunneling behavior at low temperatures, and percolation combined with activation-assisted tunneling at higher temperatures.

It should be noted that the temperature dependence of percolation conductivity is often expressed as [37] (Wu), [38] (Bhattacharya) [39] (Nenashev) [40] (Reedijk) [41] (Triberis):

$$\rho(T) = \rho_0 \times exp\left[\left(\frac{T_0}{T}\right)^x\right],$$

where the exponent $x$ depends on the system's dimensionality. Most of these studies are based on the Abeles-Sheng model [42] and its derivatives. In ideal 3D percolative systems with randomly distributed spherical particles, the exponent $x$ is typically close to 1/2 [37] (Wu). Surprisingly, in our case, the experimentally observed exponent $x \approx 1/4$ suggests an unusual correspondence between percolative behavior and 3D VRH. This can be attributed to the reduced effective dimensionality of the sample, caused by the preferential alignment of needle-like $VO_2$ nanorods within the pressing plane. The resulting anisotropic percolative network restricts current pathways, enhancing the temperature dependence of resistance and bringing the percolative exponent closer to the value observed in 3D VRH systems.

At this point, we must consider the sample's structure — a pressed metal powder in which the measuring current is unevenly distributed due to the granular morphology, the varying heights of the contact barriers between adjacent $VO_2$ nanoparticles (X-ray diffraction reveals the presence of approximately 1% of other vanadium oxides despite the high claimed purity of the $VO_2$ sample), and the anisotropy induced by uniaxial pressing. Similar effects have been observed in compacted $CrO_2$ powders, where interparticle tunneling and nonlinear transport phenomena were strongly influenced by the nature of intergranular dielectric barriers [43] [44]. These mechanisms may also play a role in the transport behavior of the $VO_2$ system.

It should be noted that percolation theory, despite its widespread use, is primarily formulated for highly idealized systems, such as uniformly distributed spherical particles with minimal variation in barrier parameters. However, real systems, with their inherent structural and compositional inhomogeneities, often exhibit transport behavior that fundamentally deviates from such simplified models. The complexity of anisotropic granular systems arises from the interplay of disorder, barrier variability, and directional anisotropy, making straightforward theoretical predictions insufficient. In the absence of a comprehensive theoretical framework that fully captures these effects, empirical analysis remains the most effective approach to understanding the transport properties of $VO_2$ powders.

Also, at low temperatures, where electron localization plays a dominant role, the percolative network may overlap with regions exhibiting VRH-like hopping between localized states near the Fermi level. As the temperature increases, the contribution from VRH diminishes, and the conductivity becomes dominated by activation-assisted hopping through the anisotropic percolative network as electrons traverse energy barriers between grains with varying heights and between adjacent conducting planes formed during uniaxial compression and subsequent relaxation [45]. The presence of residual $V_2O_5$ on the surfaces of $VO_2$ grains, possibly formed during synthesis or subsequent environmental oxidation, further enhances the percolative nature by creating insulating barriers that influence the tunneling probability and intergrain connectivity.

Thus, the inflection point observed in Mott coordinates (Fig. 6) near $T = 340$ K corresponds to the MIT in $VO_2$. However, the resistance change at this transition is significantly less pronounced than in single crystals or well-connected $VO_2$ systems. The granular structure of the pressed powder and the resulting percolative network dynamically alter intergranular connectivity during the MIT, as $VO_2$ grains expand during the metal-dielectric transition. This redistribution of current pathways within the percolation network effectively masks the expected sharp resistance jump. Although the density of states at the Fermi level $N(E_F)$ does change during the MIT, the observed inflection in the temperature dependence of resistance, plotted in Mott coordinates, is primarily governed not by changes in $N(E_F)$ due to the metal-insulator transition, but by structural reorganization within the percolative system, associated with changes in the geometric dimensions of $VO_2$ nanoparticles during the transition and the redistribution of current pathways within the composite.

In an ideal $VO_2$ system, the MIT is characterized by a resistance change of several orders of magnitude. However, in our sample, the expected dramatic resistance jump is effectively "washed out" due to several contributing factors:

- The possible presence of insulating $V_2O_5$ barriers, which limit intergranular conductivity and hinder the formation of continuous conductive pathways. These $V_2O_5$ surface layers of that may be formed on the $VO_2$ particles under the influence of the external environment [46] [47].

- Uneven shape and size distribution of $VO_2$ particles, creating anisotropic and non-uniform permeation paths and disrupting the uniform flow of current. The highly anisotropic, needle-like shape of the $VO_2$ nanoparticles with a high aspect ratio further complicates the percolation dynamics by introducing spatial inhomogeneities and changing the barrier heights. Particle compression during sample preparation leads to the preferential in-plane orientation of the grains, effectively restricting percolation pathways to fewer dimensionality and enhancing the dominance of percolation mechanisms.

- Variability in the transition temperatures across individual grains due to their small sizes and structural inhomogeneities. As seen in Fig. 6, the system's phase transition and percolative restructuring occur over a broad temperature range of 300 – 360 K.

- As a result, the system retains residual localized states even above the transition temperature and does not exhibit a fully metallic behavior. Instead, the conductivity continues to follow a temperature dependence that mimics the VRH model, although the underlying mechanism is primarily percolative.

As a result, even after the transition, the system retains residual localized states and does not exhibit a fully metallic behavior. Instead, the conductivity continues to follow a temperature dependence that mimics the VRH model, although the underlying mechanism is primarily percolative. This non-metallic behavior has been observed in other studies of thin films and even bulk $VO_2$ samples, where possible phase mixing obscures the expected change in the sign of the slope in the temperature dependence of conductivity after the transition to the metallic state.

## Conclusion

The resistance behavior in pressed $VO_2$ nanoparticle samples is dominated by percolative conductivity, with two distinct regions reflecting changes in the distribution of current paths and contact barriers before and after the metal-insulator transition (MIT). The observed agreement with Mott's variable-range hopping (VRH) at low temperatures is primarily a result of formal similarities in their mathematical descriptions, but the underlying mechanism is percolative, influenced by structural reorganization and anisotropic pathways.

During the MIT, the ~1%–2% volume contraction of $VO_2$ grains dynamically alters the interparticle tunneling barriers, leading to a redistribution of conductive paths within the percolative network. This restructuring effectively masks the expected sharp resistance jump typically observed in single-crystal $VO_2$ systems. The inflection point in the Mott coordinate plot, initially attributed to changes in the density of states $N(E_F)$, is instead linked to this structural reorganization of the conductive percolation system.

Thus, the two distinct regions in the resistance plot correspond to different percolative regimes:

- **Before the MIT ($T$ < 340 K):** Percolative transport occurs through a network of semiconducting $VO_2$ granules with a narrow bandgap (~ 0.6 – 0.7 eV), where localized states, which trap electrons due to structural disorder, and insulating barriers hinder electron flow.

- **After the MIT ($T$ > 340 K):** Although the transition to a metallic phase increases the density of states $N(E_F)$, residual barriers and structural disorder prevent the sample from exhibiting fully metallic behavior. Instead, the resistance follows a percolative exponential dependence, with activation-assisted tunneling playing a secondary role.

Unlike in classical Mott VRH conduction, where hopping occurs between localized states governed by the density of states $N(E_F)$ and the localization length ξ, in our case, the localization arises from the intergranular barriers and structural inhomogeneities. This effectively mimics the VRH temperature dependence, but the underlying mechanism is activation-assisted tunneling through a percolative network. Although the observed linear dependence $\rho(T) = \rho_0 \times exp[(T_0/T)^{1/4}]$ fits well with the Mott VRH formalism, this behavior primarily reflects the granular structure of the pressed powder. The percolative network effectively creates localized states at intergranular barriers, mimicking the VRH behavior but with underlying percolative tunneling between grains.

In conclusion, the interplay of intergranular barriers, structural inhomogeneities, and phase coexistence suppresses the expected resistive jump during the MIT, with the dynamic reorganization of current pathways within the percolative network governing the observed resistance behavior. These results underscore the critical role of percolation mechanisms in determining transport properties in granular systems like pressed $VO_2$ powders and highlight the need to account for such effects when analyzing strongly correlated materials.

## Acknowledgment

E.Yu. Beliayev gratefully acknowledges the financial support provided by the Carlsberg Foundation Denmark through the SARU (Scholars at Risk from Ukrainian Universities) Fellowship, which has enabled my work at SDU Sønderborg during this critical war situation. I also wish to express my deep gratitude to my late colleague, Nina Dalakova, who initiated this research on vanadium dioxide nanoparticles. Additionally, I am sincerely grateful to Prof. Yogendra Kumar Mishra and Prof. Horst-Günter Rubahn for their kind invitation and support, which allowed me to continue my research at Southern Denmark University, Sønderborg.